\documentclass[aip,jcp,reprint]{revtex4-1}
\usepackage{graphicx}% Include figure files

\usepackage{amsfonts, amsmath, amssymb,latexsym}

\newcommand{\dif}{\mathrm{d}}

\begin{document}

\title{Direct observation of a hydrophobic bond in
loop-closure of a capped  (-OCH$_2$CH$_2$-)$_n$ oligomer in water}

\author{Mangesh I. Chaudhari} \email{mchaudha@tulane.edu}
\author{Lawrence R. Pratt} \email{lpratt@tulane.edu}
\affiliation{Department of Chemical and Biomolecular Engineering, Tulane
University, New Orleans, LA 70118 USA} 
\author{Michael E. Paulaitis}
\email{paulaitis.1@osu.edu} 
\affiliation{Department of Chemical and
Biomolecular Engineering, Ohio State University, Columbus OH 43210 USA}

\date{\today}

\begin{abstract}
The small $r$ variation of the probability density $P(r)$ for end-to-end
separations of a {-CH$_2$CH$_3$} capped (-OCH$_2$CH$_2$-)$_n$ oligomer
in water is computed to be closely similar to the CH$_4\cdots$ CH$_4$
potential of mean force under the same circumstances. Since the aqueous
solution CH$_4\cdots$ CH$_4$ potential of mean force is the natural
physical definition of a primitive hydrophobic bond, the present result
identifies an experimentally accessible circumstance for direct
observation of a hydrophobic bond which has not been observed previously
because of the low solubility of CH$_4$  in water.  The physical picture
is that the soluble chain molecule carries the capping groups into
aqueous solution, and permits them to find one another with reasonable
frequency. Comparison with the corresponding results without the solvent
shows that hydration of the solute oxygen atoms swells the chain
molecule globule.   This supports the view that the chain molecule
globule might have a secondary effect on the hydrophobic interaction
which is of first interest here. The volume of the chain molecule
globule is important for comparing the probabilities with and without
solvent because it characterizes the local concentration of capping
groups. Study of other capping groups to enable X-ray and neutron
diffraction measurements of $P(r)$ is discussed.

\end{abstract}

\maketitle

Hydrophobic interactions are central to super-molecular self-assembly in
aqueous solutions
\cite{TANFORDC:Thehea,ISI:A1991EP16300008,Dill:1990p14722,Tanford:97,%
Harder:1998p14797,Stevens:2008p14796}, including the
folding of soluble globular proteins.\cite{Dill:2008p14719} A
surprizing development of recent years is that for entropy-dominated
hydrophobic solubilities we now have statistical mechanical theories
that are fully defensible on a molecular scale, exploiting all
available molecular-scale data.\cite{PrattLR:Molthe,AshbaughHS:Colspt,AsthagiriD.:NonWth} In contrast,
our understanding of the statistical solvent-induced forces between neighboring
small hydrocarbon molecules in water --- hydrophobic interactions ---
has \emph{not} experienced the correspondingly conclusive progress
\cite{Asthagiri:2008p1418} despite intense
\cite{WATANABEK:MOLSOT,SobolewskiE._jp070594t} computational effort. 

The principal impediment to progress in understanding hydrophobic
interactions is the lack of an accessible direct observation of a
primitive hydrophobic bond, such as the CH$_4\cdots$ CH$_4$ potential of
mean force. Because simple hydrophobic species, such as CH$_4$, are only
sparingly soluble in water,  direct observations of primitive
hydrophobic interactions are difficult to achieve; instead the influence
of hydrophobic interactions is inferred in more complex systems ---
protein folding being the foremost example.\cite{Dill:2008p14719} 
Where limited  thermodynamic evaluation of hydrophobic interactions has been
experimentally achieved for soluble, but
more complex hydrophobic species with higher
aqueous solubilities,\cite{TuckerEE:PROHI-,BernalP:VAPSOH} 
those experiments set-off a substantial modeling effort that has not
untangled this complicated issue.\cite{Asthagiri:2008p1418}

\begin{figure}[h]
\includegraphics[width=3.2in]{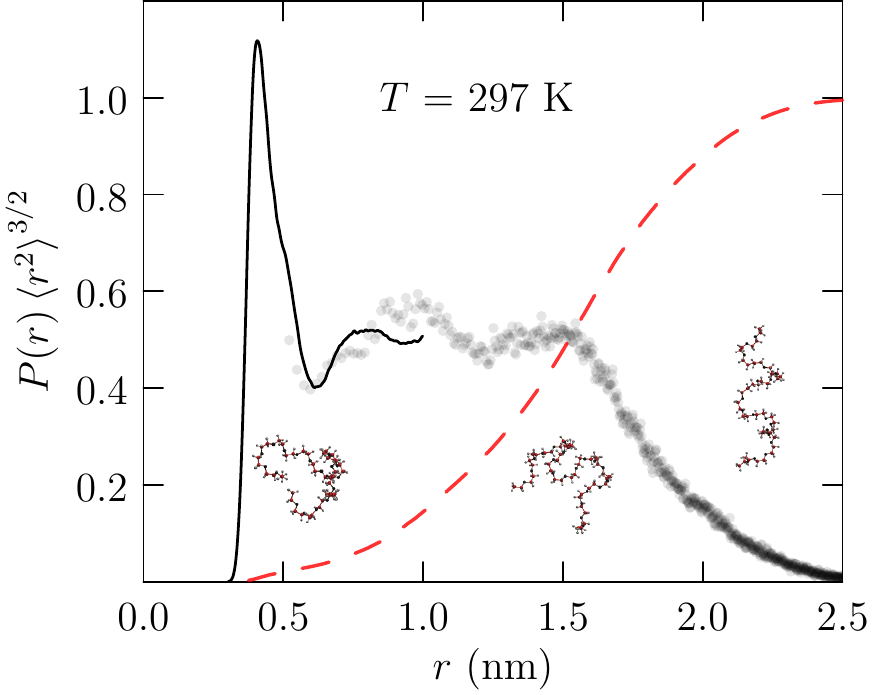}
\caption{Probability density $P(r)$ for radial
displacement of end-C atoms of
CH$_3$CH$_2$O(CH$_2$CH$_2$O)$_{20}$CH$_2$CH$_3$ in aqueous solution,
with the cumulative distribution $ \int_0^r P(x) 4\pi {x}^2 \dif x$
(dashed line). The dots were obtained by parallel tempering \cite{Earl:2005p11896} over times
of 20~ns/replica with 32 temperatures spanning 256-550~K, using AMBER
10 \cite{amber} with the same molecule numbers and cubical volume $V =
32.254$~nm$^3$ for each replica. Monte Carlo swaps of configurations had
an overall success rate of 23\%. The SPC/E model treated the 1000 water
molecules \cite{BEREND87A}, and an extended atom model with the
Generalized Amber Force Field (GAFF) was used for the single chain
molecule.\cite{gaff} The solid line is a stratified recalculation of the
loop-closure feature, adopting a harmonic potential for the radial
displacement coordinate covering the range of 0.3-1.0 nm uniformly with
15 windows, then reconstituting $P(r)$ on that range with a weighted
histogram method.\cite{FrenkelSmit,AlanGrossfield,Shell}   This second set of
calculations lasted 20~ns/window. The undetermined multiplicative
constant in the stratified calculation of $P(r)$ is adjusted to match
the direct observation from the parallel tempering in the region of the
first minimum.  The left vertical axis is non-dimensionalized with the
observed $\left\langle r^2\right\rangle{}^{1/2}$ = 1.56~nm.}
\label{fig:PR}
\end{figure}

Here we show (FIG.~\ref{fig:PR}) that the probability density $P(r)$ of
end-to-end separations of a {-CH$_2$CH$_3$} capped (-OCH$_2$CH$_2$-)$_n$
oligomer in water exhibits  a distinct hydrophobic bond between the two
end C-atoms. These atom-pair correlations are intrinsically measurable
by X-ray and neutron diffraction.\cite{Annis:2001p15716}  A natural
view of the present case is that the chain molecule carries this
hydrophobic pair into solution and permits them to find one another with
reasonable frequency. The underlying assumption is that the hydrophobic
interaction is sufficiently local that the effects of the supporting
chain molecule are secondary.   This is a reasonable assumption that can
be experimentally tested by variation of the chain length and capping
groups.  Nonetheless, the present realization is consistent with the
view that hydrophobic interactions are typically expressed in the
context of other effects in micelles, membranes, and the structure of
soluble proteins.

\begin{figure}
\includegraphics[width=3.2in]{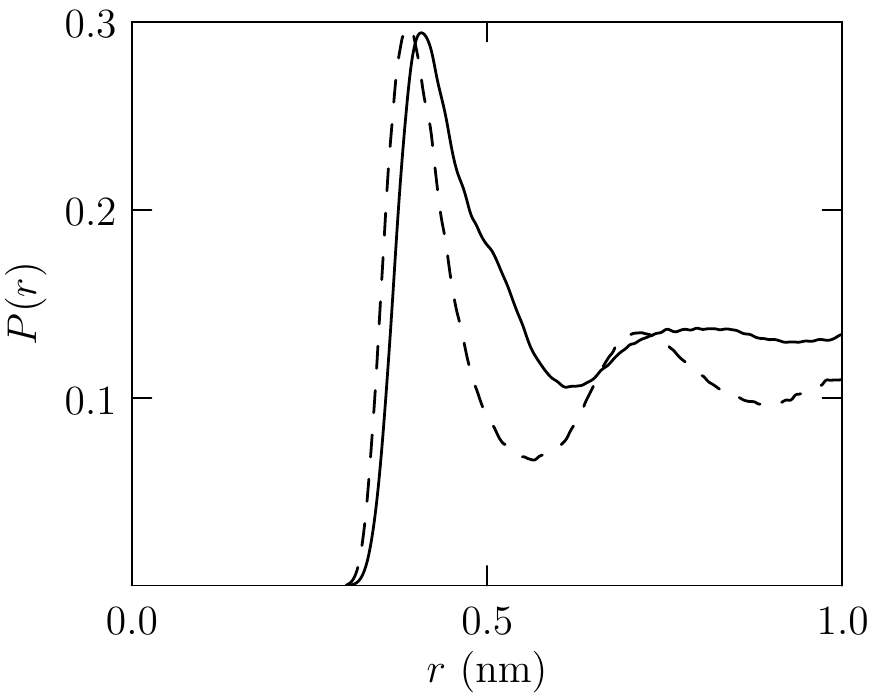} \caption{Comparison of the
probability density of FIG.~\ref{fig:PR} (solid curve) with results from the
CH$_4\cdots$ CH$_4$ potential of mean force obtained with the same
models and stratification methods (dashed curve) as FIG.~\ref{fig:PR}.  The
latter result is in good agreement with a previous simulation evaluation
for a similar model.\cite{SobolewskiE._jp070594t}  The multiplicative
constant required for this comparison was adjusted informally for
agreement in the region of the maxima of these two results.}
\label{fig:gor}
\end{figure}

Two distinct sets of calculations were combined to obtain the present
results:   firstly, parallel tempering \cite{Earl:2005p11896} to establish the overall
structure of $P(r)$ and, secondly, windowing \cite{Shell} to achieve satisfactory
spatial resolution in the interesting loop-closure regime, $r <
0.7$~nm.  Direct observation of the contact feature
would rest on a few percent of the parallel-tempering data set,
\emph{i.e.,} about 2\% of the molecules are in loop-closure
configurations. The overall parallel-tempering results permit evaluation of the
undetermined scale factor (zero of the potential of the average forces)
for the windowing results.

The shape of $P(r)$ in the loop-closure region is strikingly similar to
predicted pair distributions for model inert gases in water.\cite{ISI:A1986E409400003,Trzesniak:2007p14844,Asthagiri:2008p1418} On
the basis of the stratified evaluation of $P(r)$, the most probable
--CH$_3 \cdots$ CH$_3$-- hydrophobic bond length is about 0.4~nm. The
slight shoulder on the large-$r$ side of the principal maximum of $P(r)$
is a remnant of a RISM cusp \cite{Ladanyi:1975p14723} and reflects the
ethyl end-capping.

A direct comparison  with results for CH$_4\cdots$ CH$_4$ obtained with
the same models and stratification methods (FIG.~\ref{fig:gor}) shows good agreement in the placement of the principal peak, and the
large-$r$ shoulder of the principal peak is more obvious.  Contact
pairing configurations are more prominent than solvent-separated ones in
both of these cases, supporting a standard view of this hydrophobic
interaction.  A separate evaluation (not shown) of the solubility
of CH$_4$(aq) with these models compares well with experiment. This
substantiates the view that this model solute naturally represents
hydrophobic methane. These probability densities differ in outer shells
due to detailed differences between the solvent and the chain molecule
medium.

\begin{figure}
\includegraphics[width=3.2in]{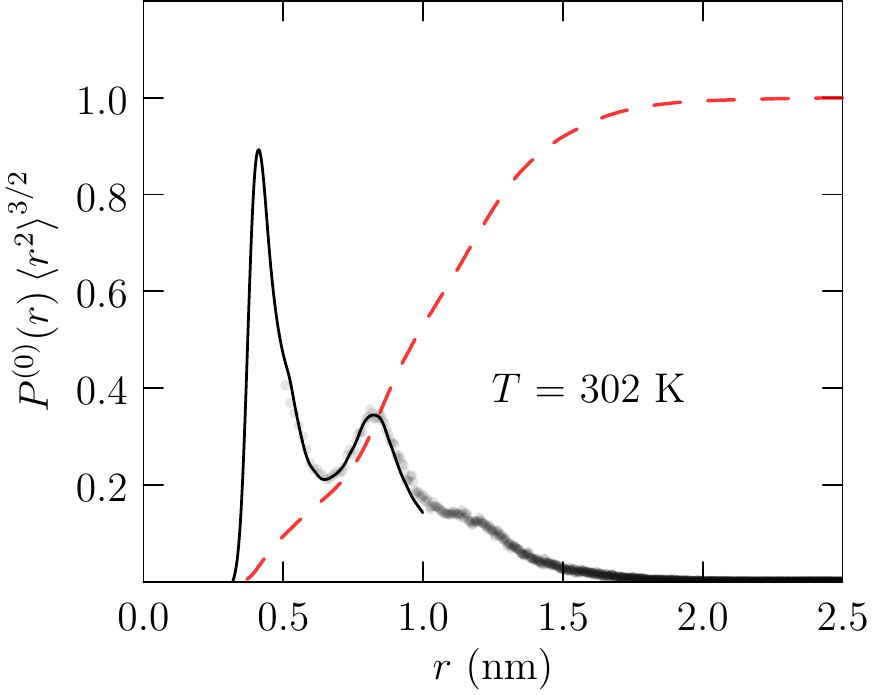} \caption{As in
FIG.~\ref{fig:PR} but without water present. The left vertical axis is
non-dimensionalized with the observed $\left\langle
r^2\right\rangle{}^{1/2}$ = 1.06~nm.}
\label{fig:nosolvent}
\end{figure}

It is natural also to compare these pdfs with what would be obtained if
the solvent were absent (FIG.~\ref{fig:nosolvent}). The comparison shows that
water swells this soluble chain molecule (FIG.~\ref{fig:PR}), as is the case
also with aqueous dilution of a PEO melt.\cite{Smith:2000p15718} This
supports the view that the chain molecule globule might have a secondary
effect on the hydrophobic interaction which is of first interest here.

Note that the boundary between the low-extension $r < \left\langle
r^2\right\rangle{}^{1/2}$  region and a high-extension region  is
definite (FIG.~\ref{fig:PR}). On the basis of the similarity with the
results for FIG.~\ref{fig:gor}, the end-cap pair appears to be shielded
within a
uniform fluid environment in the low-extension region.
Furthermore, end-caps are more concentrated in the
smaller-volume polymer globule of FIG.~\ref{fig:nosolvent} than in the
larger-volume polymer globule of FIG.~\ref{fig:PR}. The
non-dimensionalized probability densities incorporate that distinction
without which the maximum probabilities would differ by nearly a factor
of three. This point is consistent with the observation from
FIG.~\ref{fig:gor} that if the probability densities there match roughly
at large-$r$, then the maximum values also approximately match.

The high-extension $r > \left\langle r^2\right\rangle{}^{1/2}$ tail of
the probability density (FIG.~\ref{fig:PR}) is reasonably described by a
linear-response behavior
\begin{eqnarray}
-\frac{\dif}{\dif r} \ln P(r)\approx a + b r~.
\end{eqnarray}
A Gaussian model for $P(r)$ at high-extension  would not be centered on the origin,
however, and in that sense a traditional Gaussian model would be
unsatisfactory in the high-extension region.

Simple characterizations of hydrophobic effects, \emph{i.e.} whether
they are attractive or repulsive,  depend on the specific properties
examined and the comparisons made. For example, well-developed
theoretical analyses \cite{WATANABEK:MOLSOT} show that hydrophobic
interactions can be repulsive for the osmotic second virial coefficient.
 Other comparisons illuminate different aspects of hydrophobic
interactions. Sometimes hydrophobic interactions are judged by adopting
another solvent that provides a natural comparison to the case of the
water medium.\cite{Nozaki:1971p15584}  Sometimes it is most direct to
judge hydrophobic interactions by comparison with standardly hydrophilic
solutes, polar, H-bonding,  or ionic species in water.  We propose that
identification of an experimentally accessible case that also permits
detailed molecular theory and computation should assist in resolving
such alternatives.

Though this hydrophobic interaction is intrinsically measurable,
analysis of X-ray and neutron diffraction experiments on such systems
will require specific and thorough support from molecular simulations,
as is the current practice.\cite{Annis:2001p15716,Soper:2007p14783}
Isotopic substitution and labeling will be essential. Consideration of
fluorinated caps such as --CF$_3$
\cite{Bonnet:2007p14799,Bonnet:2010p14798} should make this hydrophobic
interaction more prominent yet.\cite{AsthagiriD.:NonWth}

The chain molecules considered here have broad technological interest
because of their biocompatibilities.\cite{Jevsevar:2010p15716} They are
also intrinsic to the dispersant materials used in response to oil
spills.\cite{dispersants} Small angle neutron scattering
\cite{Hammouda:2004p14654} and fluorescence and light-scattering studies
\cite{Alami:1997p15714} have shown that the aqueous solution
interactions of (-OCH$_2$CH$_2$-)$_n$ polymers are sensitive to the
end-capping of the chains.  The importance of end-effects is also
supported by the sensitivity of solution phase diagrams to the
(-OCH$_2$CH$_2$-)$_n$ lengths.\cite{Dormidontova:2004p14819,Cote:2008p14817} These observations
suggest hydrophobic bonding of the -CH$_3$ caps, and that manipulation
of capping groups might help in understanding hydrophobic interactions
on a molecular scale \cite{PrattLR:Molthe} by exhibiting a localized
hydrophobic bond.

We thank D. Asthagiri and M. Berkowitz for helpful discussions.

%\bibliography{ECapPairing} 


\end{document}